\documentstyle[epsfig,aps]{revtex}
\begin{document}
\draft
\def\lsim{\lower.5ex\hbox{$\; \buildrel < \over \sim \;$}}
\def\gsim{\lower.5ex\hbox{$\; \buildrel > \over \sim \;$}}
\title{ Dynamics of electromagnetic waves in Kerr geometry}
\author  {Banibrata Mukhopadhyay\\}
\address{Theoretical Physics Group,\\
Physical Research Laboratory,\\
Navrangpura, Ahmedabad-380009, India\\}

\thanks{e-mail:  bm@prl.ernet.in } \\
\maketitle
\baselineskip = 18 true pt
\vskip0.3cm
\setcounter{page}{1}
\def\ch{\lower-0.55ex\hbox{--}\kern-0.55em{\lower0.15ex\hbox{$h$}}}
\def\lh{\lower-0.55ex\hbox{--}\kern-0.55em{\lower0.15ex\hbox{$\lambda$}}}	

\begin{abstract}

Here we are interested to study the spin-1 particle i.e., electro-magnetic wave
in curved space-time, say around black hole. After separating
the equations into radial and angular parts, writing them according to
the black hole geometry, say, Kerr black hole we solve them analytically.
Finally we produce complete solution of the spin-1 particles around
a rotating black hole namely in Kerr geometry. Obviously there is coupling
between spin of the electro-magnetic wave and that of black hole when
particles propagate in that space-time. So the solution will be
depending on that coupling strength. This solution may be useful to 
study different other problems where the analytical results are needed. 
Also the results may be useful in some astrophysical contexts.

\end{abstract}

\vskip1.0cm
\vskip2cm
\section*{I. INTRODUCTION}
We know that Maxwell's equations describe the dynamics of
electro-magnetic wave. Four equations indicate how electric and magnetic
field are dependent each other and how they propagate in space.
Also electro-magnetic wave means particle of spin-1, like photon. 
To study the dynamics of the spin-1 particles in curved space-time 
one needs to look up the Maxwell's equation in Kerr geometry (if the
central gravitating object is chosen rotating and uncharged). In this
context it is easier to consider the electromagnetic waves
as an external perturbation in the space-time. This external perturbation
can be represented as incident waves of different sorts in the space-time
of the black hole. We are considering the back ground space-time 
as of Kerr black hole which is stationary and axisymmetric. So the perturbation
can be expressed as a superposition of waves with different modes with the 
`$t$' and `$\phi$' dependence given as $exp[i(\sigma t+m\phi)]$, where
$\sigma$ is the frequency of the waves and $m$ is an integer which may
be positive, negative or zero. Far away from the black hole the space-time
is flat where the Maxwell's equation and its solution are well known.
As the electromagnetic wave i.e., corresponding perturbation comes
closer to the black hole the space-time becomes curved where 
usual quantum theory may not be applicable. In that region, set of Maxwell
equations will be modified and corresponding global behaviour of the
spin-1 particles will be deviated with respect
to that of flat space.

Photon is an electro-magnetic radiation. From the study of Maxwell equations
we can investigate the behaviour of the photon close to the black hole
or other gravitating object which result can be applicable further 
for other related works and in the context
of astrophysical problem where the relativistic information  
is important. The solutions of Maxwell equations in a curved space-time
have a considerable importance in black hole physics as well as in the context
of mathematical physics. If the spatially complete solution is possible
to construct, it is very useful to study the absorption rate of the black 
hole, to find the corresponding Feynman Green function,
for checking the stability of the black hole (say, in case of the
Kerr geometry, one can study the stability of Kerr black hole), to study
the black hole perturbation, to study the quasi-normal modes of the black hole,
to do the second quantization of the metric (in this case Kerr metric) etc.
In case of matter flows towards the compact object,
out of the matter which falls towards the central body photon is there.
That photon also may take part in the cooling process of the accreting matter,
though the Compton wavelength of the incoming photon is very small compared 
to size of the black hole horizon if the black hole is chosen as
formed by the supernova explosion only. On the other
hard, we know all the primordial black holes of mass greater than
$10^{15}$gm are still exist in nature\cite{st}. For this kind of
black holes the interaction between the incoming photon and the black hole
may be significant. Thus, naturally the dynamical behaviour of the photon, how it is coupled with
space-time (how the spin of the photon couples with the rotation of  
compact object) is very important to study. Also, in the case of Hawking
radiation, boson, fermion, graviton are emitted from the black hole
and scattered in the space-time. In that context the behaviour of the
electromagnetic wave is essential to look into.

In 1972, Teukolsky\cite{teu72} initiated the electromagnetic perturbation in 
Kerr geometry and separated corresponding master equation. It was
shown that, separation into radial ($r$) and angular ($\theta$) parts
is possible if the particular choice of temporal ($t$) and azimuthal
($\phi$) dependency is chosen in case of stationary and axisymmetric
system as given above. In next year Teukolsky\cite{teu73} himself
studied corresponding reflection and transmission problem of the 
electromagnetic waves in Kerr geometry. Then, Teukolsky \& Press\cite{tp}
studied the interaction of electromagnetic waves with black hole 
and found numerical solutions asymptotically. In 1976, 
Chandrasekhar\cite{chandra76} again looked into the separated Maxwell equations
in curved space-time particularly in Kerr geometry and indicated about some
features like super-radiance. Later he discussed\cite{chandra83} 
extensively about the electromagnetic, Dirac, gravitational wave equation 
in curved space-time and showed some asymptotic results. In 1984, 
Chakrabarti\cite{chakra} solved the separated angular Dirac equation 
analytically and also listed the values of the separation constant $\lh$
for the particle of spin 1/2, 1 and 2. Then, various scattering phenomena
from black holes have been studied by Futterman et al.\cite{f88}.
After that, recently,
Mukhopadhyay \& Chakrabarti\cite{mc99,mc00,cm00},
Mukhopadhyay\cite{m99,m00} have started to study different behaviour
of spin-1/2 particles in curved space-time in case of non-rotating 
(Schwarzschild), rotating (Kerr) and charged (Reissner-Nordstr\"om) 
black hole geometry. Following (almost) same approach here we would
like to re-initiate the study of spin-1 particles in Kerr geometry.
Most of the earlier studies\cite{teu73,tp,chandra76} have given asymptotic
solutions of Maxwell equations, here we are going to present spatially
complete, global solution in Kerr geometry. Here one of the main difference
of our approach is that after separating the equation into radial and angular part,
with a further certain choice of basis, imaginary number $i$
($=\sqrt{-1}$) is completely taken off from potential of the system. Physically, as if
we consider such a frame there the potential is real which is felt by the 
incoming electromagnetic wave. So the final equations to be studied 
are explicitly real which makes easier to attack the problem (in case
of \cite{chandra83}, potential contains real and imaginary both the parts,
so the potential behaviour had to be studied separately).

We will use Newman-Penrose
spinor formalism to write the Maxwell equations in curved space-time and then
separate into the radial and angular part to solve separately. In the next
section we will introduce the basic equations of the problem. In \S 3, we will
show analytical solutions where the behaviour of potentials for the different sets of input
parameter are shown. Finally, in \S 4, we will make conclusions.

\section*{II. Basic Equations of the Problem}

We know that $F_{ij}=\partial_i A_j-\partial_j A_i$ is the electromagnetic 
covariant antisymmetric tensor where, $A_i=(A_0,{\overrightarrow A})$ 
($A_0$ and ${\overrightarrow A}$ are the scalar and vector potential respectively) 
and corresponding Maxwell equations in curved space-time is 
\begin{equation}
\label{max1}
F_{[ij;k]}=0,\hskip1cm g^{ik}F_{ij;k}=0.
\end{equation}
Following Chandrasekhar\cite{chandra83}, using Newman-Penrose 
formalism, $F_{ij}$ can be written in terms of three complex scalars as
\begin{eqnarray}
\label{max2}
\phi_0=F_{13} & =&  F_{ij}l^i m^j,\nonumber \\
\phi_1=\frac{1}{2}(F_{12}+F_{43})&  =& \frac{1}{2}F_{ij}(l^i n^j+{\bar m}^i m^j),\\
\phi_2=F_{42}&  = & F_{ij}{\bar m}^i m^j.\nonumber
\end{eqnarray}
Here, $l^i$, $n^i$, $m^i$ and ${\bar m}^i$ are four null basis for 
Newman-Penrose formalism. In terms of the tetrad components and intrinsic 
derivatives Maxwell equations become
\begin{eqnarray}
\label{max3}
F_{[ab|c]}=0,\hskip1cm \eta^{nm}F_{an|m}=0.
\end{eqnarray}
Thus, by using eqns. (\ref{max2}) and (\ref{max3}), Maxwell equations are 
replaced by
\begin{eqnarray}
\label{max4}
\phi_{1|1}-\phi_{0|4}=0, \hskip1cm\phi_{2|1}-\phi_{1|4}=0,\\
\phi_{1|3}-\phi_{0|2}=0, \hskip1cm\phi_{2|3}-\phi_{1|2}=0.\nonumber
\end{eqnarray}
In terms of different spin coefficients, tetrads which are expressed
in terms of directional derivatives, eqn. (\ref{max4}) 
i.e., the reduced Maxwell equations become
\begin{eqnarray}
\label{max5}
D\phi_1-\delta^*\phi_0 &=& (\pi-2\alpha)\phi_0+2\rho\phi_1-\kappa\phi_2, \nonumber \\
D\phi_2-\delta^*\phi_1 &=& -\lambda\phi_0+2\pi\phi_1+(\rho-2\epsilon)\phi_2,\\
\delta\phi_1-{\underline \Delta}\phi_0&=&(\mu-2\gamma)\phi_0+2\tau\phi_1
-\sigma\phi_2, \nonumber \\
\delta\phi_2-{\underline \Delta}\phi_1&=&-\nu\phi_0+2\mu\phi_1+(\tau-2\beta)\phi_2.\nonumber
\end{eqnarray}
Here, $D$, ${\underline \Delta}$, $\delta$, $\delta^*$ are the directional
derivatives which are basically the basis vectors of the system. Different spin
coefficients namely $\pi$, $\alpha$, $\rho$, $\kappa$, $\lambda$, $\epsilon$,
$\mu$, $\gamma$, $\tau$, $\sigma$, $\nu$, $\beta$ are given by 
Chandrasekhar\cite{chandra83} and which are expressed in terms of the Kerr 
metric coefficients. If we substitute all these in
eqn. (\ref{max5}) we get 
\begin{eqnarray}
\label{max6}
\frac{1}{\bar{\rho}^*\sqrt{2}}\left({\cal L}_1-\frac{iasin\theta}
{\bar{\rho}^*}\right)\phi_0 &=&\left({\cal D}_1+\frac{2}
{\bar{\rho}^*}\right)\phi_1, \nonumber \\
\frac{1}{\bar{\rho}^*\sqrt{2}}\left({\cal L}_0+\frac{2iasin\theta}
{\bar{\rho}^*}\right)\phi_1 &=&\left({\cal D}_0+\frac{1}
{\bar{\rho}^*}\right)\phi_2,  \\
\frac{1}{\bar{\rho}\sqrt{2}}\left({\cal L}_1^\dagger+\frac{iasin\theta}
{\bar{\rho}^*}\right)\phi_2 &=&-\frac{\Delta}{2\rho^2}\left({\cal D}_0^\dagger
+\frac{2}{\bar{\rho}^*}\right)\phi_1, \nonumber \\
\frac{1}{\bar{\rho}\sqrt{2}}\left({\cal L}_0^\dagger+\frac{2iasin\theta}
{\bar{\rho}^*}\right)\phi_1 &=&-\frac{\Delta}{2\rho^2}\left({\cal D}_1^\dagger
-\frac{1}{\bar{\rho}^*}\right)\phi_0, \nonumber 
\end{eqnarray}
where, ${\bar \rho}=r+ia cos\theta$, ${\bar \rho}^*=r-ia cos\theta$ and 
${\cal L}_n$,  ${\cal L}_n^\dagger$, ${\cal D}_n$ and ${\cal D}_n^\dagger$
are defined as
$$
{\cal D}_n=\partial_r+\frac{iK}{\Delta}+2n\frac{r-M}{\Delta};\hskip0.5cm
{\cal D}_n^\dagger=\partial_r-\frac{iK}{\Delta}+2n\frac{r-M}{\Delta},
\eqno{(7a)}
$$
$$
{\cal L}_n=\partial_\theta+Q+ncot\theta;\hskip0.5cm
{\cal L}_n^\dagger=\partial_\theta-Q+ncot\theta, 
\eqno{(7b)}
$$
$$
K=(r^2+a^2)\sigma+am;\hskip0.2cm \Delta=r^2+a^2-2Mr; \hskip0.2cm 
Q=a\sigma sin\theta+mcosec\theta.
\eqno{(7c)}
$$
Now choosing $\Phi_0=\phi_0$, $\Phi_1=\phi_1{\bar \rho}^*\sqrt{2}$ and
$\Phi_2=2\phi_2({\bar \rho}^*)^2$ eqn. (\ref{max6}) reduces to
\setcounter{equation}{7}
\begin{eqnarray}
\label{max7}
\left({\cal L}_1-\frac{iasin\theta}{\bar{\rho}^*}\right)\Phi_0&=&
\left({\cal D}_1+\frac{1}{\bar{\rho}^*}\right)\Phi_1, \nonumber \\
\left({\cal L}_0+\frac{iasin\theta}{\bar{\rho}^*}\right)\Phi_1&=&
\left({\cal D}_0-\frac{1}{\bar{\rho}^*}\right)\Phi_2, \\
\left({\cal L}_1^\dagger-\frac{iasin\theta}{\bar{\rho}^*}\right)\Phi_2 &=&
-\Delta\left({\cal D}_0^\dagger+\frac{1}{\bar{\rho}^*}\right)\Phi_1, \nonumber \\
\left({\cal L}_0^\dagger+\frac{iasin\theta}{\bar{\rho}^*}\right)\Phi_1 &=&
-\Delta\left({\cal D}_1^\dagger-\frac{1}{\bar{\rho}^*}\right)\Phi_0. \nonumber 
\end{eqnarray}
Now using first and last equation of (\ref{max7}) we get
\begin{eqnarray}
\label{radang1}
\left[\Delta{\cal D}_1{\cal D}_1^\dagger+{\cal L}_0^\dagger{\cal L}_1
-2i\sigma(r+iacos\theta)\right]\Phi_0=0
\end{eqnarray}
and using second and third of (\ref{max7}) we get  
\begin{eqnarray}
\label{radang2}
\left[\Delta{\cal D}_0^\dagger{\cal D}_0+{\cal L}_0{\cal L}_1^\dagger
+2i\sigma(r+iacos\theta)\right]\Phi_2=0.
\end{eqnarray}
Equations (\ref{radang1}) and (\ref{radang2}) are clearly separable.
Now we choose, $\Phi_0=R_{+1}(r)S_{+1}(\theta)$, $\Phi_2=R_{-1}(r)S_{-1}(\theta)$
and separating the eqns. (\ref{radang1}) and (\ref{radang2}) into radial
and angular parts we get the Teukolsky's equation as
\begin{eqnarray}
\label{radangf}
\left(\Delta{\cal D}_0{\cal D}_0^\dagger-2i\sigma r\right)\Delta R_{+1}
&=&\lh\Delta R_{+1},\nonumber\\
\left({\cal L}_0^\dagger{\cal L}_1+2a\sigma cos\theta\right)S_{+1}
&=&-\lh S_{+1},\\
\left(\Delta{\cal D}_0^\dagger{\cal D}_0+2i\sigma r\right) R_{-1}
&=&\lh R_{-1},\nonumber\\
\left({\cal L}_0{\cal L}_1^\dagger-2a\sigma cos\theta\right)S_{-1}
&=&-\lh S_{-1}.\nonumber
\end{eqnarray}
Here, we use the identity $\Delta {\cal D}_1 {\cal D}_1^\dagger R_{+1}
={\cal D}_0 {\cal D}_0^\dagger \Delta R_{+1}$ and $\lh$ is chosen as the separation
constant.
Now choosing $P_{+1}=\Delta R_{+1}$ and $P_{-1}=R_{-1}$ and 
by a suitable choice of the relative 
normalisation of the functions $P_{+1}$ and $P_{-1}$ we get
\begin{eqnarray}
\label{norf}
\Delta {\cal D}_0{\cal D}_0 P_{-1}=\xi P_{+1};\hskip0.5cm
\Delta {\cal D}_0^\dagger {\cal D}_0^\dagger P_{+1}=\xi^* P_{-1},
\end{eqnarray}
where, $|\xi|^2=\lh^2-4\alpha^2\sigma^2$ and $\alpha^2=a^2+\frac{am}{\sigma}$.
Thus, from the first of eqn. (\ref{norf}), (7a,b) and third of (\ref{radangf}) we get
\begin{eqnarray}
\label{peq1}
\xi P_{+1}=(\lh-2i\sigma r)P_{-1}+2iK{\cal D}_0 P_{-1}.
\end{eqnarray}
Similarly, using second of eqn. (\ref{norf}), (7a,b) and first of (\ref{radangf}) 
we get
\begin{eqnarray}
\label{peq2}
\xi^* P_{-1}=(\lh+2i\sigma r)P_{+1}-2iK{\cal D}_0^\dagger P_{+1}.
\end{eqnarray}
Then from eqns. (\ref{peq1}) and (\ref{peq2}) we get
\begin{eqnarray}
\label{rad1}
\frac{d}{dr}P_{+1}=\frac{iK}{\Delta}P_{+1}-\frac{i}{2K}\left[(\lh+2i\sigma r)
P_{+1}-\xi^* P_{-1}\right],
\end{eqnarray}
\begin{eqnarray}
\label{rad2}
\frac{d}{dr}P_{-1}=-\frac{iK}{\Delta}P_{-1}+\frac{i}{2K}\left[(\lh-2i\sigma r)
P_{-1}-\xi P_{+1}\right].
\end{eqnarray}
Now, eqns. (\ref{rad1}) and (\ref{rad2}) are the set of radial equations which we
will solve for the radial behaviour of spin-1 particles.

\subsection*{II.A. Reduction of Angular equations}
Now, in angular set of equation of (\ref{radangf}), if we choose the 
integrating factor $sin\theta$ and consider the independent variable
$\theta$ transforms as $\frac{d}{dU}=sin\theta\frac{d}{d\theta}$ such that
\begin{eqnarray}
\label{angtran}
U=log\left|tan\frac{\theta}{2}\right|,
\end{eqnarray}
we get the transformed set of angular Maxwell equations as
\begin{eqnarray}
\label{angtreq}
\frac{d^2 S_{\pm 1}}{dU^2}+{\cal J}^2_{\pm 1} S_{\pm 1}=0,
\end{eqnarray}
where, ${\cal J}^2_{\pm 1}=sin^2\theta (\lh\pm Q^\prime -cosec^2\theta
\mp Q cot\theta-Q^2\pm 2a\sigma cos\theta)$, prime denotes the derivative with
respect to $\theta$.

According to the transformation of variable $\theta$ (which runs from $0$ to
$\pi$) to $U$ (which runs from $-\infty$ to $+\infty$) the reduced 
eqn. (\ref{angtreq}) is now wave equation in cartesian like coordinate
system with wave vector ${\cal J}_{\pm 1}$ which contains the information of
angular potential felt by the spin-up and spin-down incoming particles.  

\subsection*{II.B. Reduction of Radial equations}
We define the new set of basis for spin-1 particles as $P_\pm=P_{+1}\pm iP_{-1}$.
So from eqns. (\ref{rad1}) and (\ref{rad2}) we get
\begin{eqnarray}
\label{rad1mod1}
\left[\frac{d}{dr}-\left(\frac{\sigma r}{K}+\frac{\xi}{2K}\right)\right]P_{+}
=\left[\frac{K}{\Delta}-\frac{\lh}{2K}\right]P_- e^{i\pi/2},
\end{eqnarray}
where $\xi$ is real\cite{chandra83} and given as $\xi=\xi^*=\sqrt{\lh^2-4\alpha^2\sigma^2}$.
Further, defining $\psi_\pm=P_\pm e^{\mp i\pi/4}$ eqn. (\ref{rad1mod1})
reduces to
\begin{eqnarray}
\label{rad1mod2}
\left[\frac{d}{dr}-P_1(r)\right]\psi_+=-W\psi_-.
\end{eqnarray}
Similarly, we can get another one as,
\begin{eqnarray}
\label{rad2mod2}
\left[\frac{d}{dr}-P_2(r)\right]\psi_-=W\psi_+.
\end{eqnarray}
Here, $P_1(r)=\frac{\sigma r}{K}+\frac{\xi}{2 K}$,
$P_2(r)=\frac{\sigma r}{K}-\frac{\xi}{2 K}$ and
$W=\frac{\lh}{2K}-\frac{K}{\Delta}$.
Now, eqns. (\ref{rad1mod2}) and (\ref{rad2mod2}) are the coupled 
radial equations for spin-1 particles in the particular choice of
basis system as given above. To get appropriate solution we will
decouple those and change the independent variable $r$ to $V$ as 
is given below.

Decoupling eqns. (\ref{rad1mod2}) and (\ref{rad2mod2}) for $\psi_+$ we get
\begin{eqnarray}
\label{rad1moddec}
\frac{d^2\psi_+}{dr^2}-\left(P_1(r)+P_2(r)+\frac{W^\prime}{W}\right)
\frac{d\psi_+}{dr}+\left(W^2-P_1^\prime+\frac{W^\prime}{W}P_1+
P_1 P_2\right)\psi_+=0,
\end{eqnarray}
where, prime denotes the derivative with respect to $r$.
Now, we consider the integrating factor $[W(r^2+\alpha^2)]^{-1}$ and variable
transformation as $\frac{d}{dV}=[W(r^2+\alpha^2)]^{-1}\frac{d}{dr}$ such that
\begin{eqnarray}
\label{radtr}
V=\left(\frac{\lh}{2\sigma}\right)r-\sigma\left[\frac{r^3}{3}+Mr^2+r(2\alpha^2
+2(2M^2-a^2)+a^2)-\frac{(\alpha^2+r_-^2)^2}{2\sqrt{M^2-a^2}}log(r-r_-)+
\frac{(\alpha^2+r_+^2)^2}{2\sqrt{M^2-a^2}}log(r-r_+)\right],
\end{eqnarray}
where, $r_\pm=M\pm\sqrt{M^2-a^2}$ and $\alpha^2=a^2+am/\sigma$.
Thus from eqn. (\ref{rad1moddec}) we get
\begin{eqnarray}
\label{rad1dec}
\frac{d^2\psi_+}{dV^2}+{\cal K}_+^2\psi_+=0,
\end{eqnarray}
where, ${\cal K}_+^2=\frac{\left(W^2-P_1^\prime+\frac{W^\prime}{W}P_1+P_1 P_2
\right)}{W^2(r^2+\alpha^2)^2}$.
Similarly, decoupling eqns. (\ref{rad1mod2}) and (\ref{rad2mod2}) for 
$\psi_-$ we get 
\begin{eqnarray}
\label{rad2dec}
\frac{d^2\psi_-}{dV^2}+{\cal K}_-^2\psi_-=0,
\end{eqnarray}
where, ${\cal K}_-^2=\frac{\left(W^2-P_2^\prime+\frac{W^\prime}{W}P_2+P_1 P_2
\right)}{W^2(r^2+\alpha^2)^2}$.
Finally, we have set of second order differential wave equations (\ref{rad1dec})
and (\ref{rad2dec}) which are easier to attack to study the radial behaviour
of spin-1 particles. Usually new transformed variable $V$ varies from $-\infty$
to $+\infty$. But, if the frequency $\sigma$ of the incident wave is such that
\begin{eqnarray}
\label{potdiv}
W^2(r^2+\alpha^2)^2=0 
\end{eqnarray}
at $r=r_{c}> r_+$, $V-r$ relation becomes multivalued. 
For that frequency range wave vector ${\cal K}_\pm$ diverges at $r=r_{c}$ outside
the horizon and the energy extraction from the space-time may be possible
in the range $r_+$ to $r_c$. Here $r_c$ may play the same role as the
radius of ergosphere. We know that, in case of a rotating black hole energy extraction
may be possible from the ergo-region. Chandrasekhar\cite{chandra83} conjectured from the
asymptotic solution of the electromagnetic wave that the super-radiance is exist.
He showed that, if the frequency of the incident wave is less than of certain cut-off
value (i.e., $\lsim -\frac{am}{2Mr_+}$), potential barrier of the 
system diverges and it varies as $\frac{1}{(r-|\alpha|)^4}$ 
close to the singular point. Here, our calculations tally well with that conjecture. 
Equations (\ref{rad1dec}) and (\ref{rad2dec}) are the
wave equations in cartesian like coordinate system where ${\cal K}_\pm$ contains
the information of the radial potential felt by the particles in the new
basis system.

Now, from eqn. (\ref{potdiv}) we see the singular points are, where potential
diverges for
$$
\sigma=\sigma_c=\frac{\pm\Delta^{1/2}\sqrt{2\lh}-2am}{2(a^2+r^2)},
\eqno{(27a)}
$$
$$
\sigma=\sigma_c=-\frac{am}{r^2+a^2}.
\eqno{(27b)}
$$
Thus, there are three possible frequencies where the potential may 
diverge and the super-radiance occurs. But not necessarily for all three $\sigma_c$s 
potential will diverge
at $r\ge r_+$ for particular $a$, $m$, $M$. If the frequencies of the incident
electromagnetic wave are such that, at a radius $r\ge r_+$ at least one of the
eqns. (27a,b) satisfy then for that frequency range super-radiance is
expected. For the other values of $\sigma$ the potential as well as 
${\cal K}_\pm^2$ behaviour does not have any singular point as is shown 
in Figs. 1 and 2a,b below. Here, the frequencies for which super-radiance
is expected to occur are looking different from that
given by Chandrasekhar\cite{chandra83}. Actually, as we change the
dependent variable ($R_{\pm 1}$ to $\psi_\pm$) and independent variable ($r$ to $V$)
here in a different manner with respect to that of Chandrasekhar, frequency
expressions are looking different apparently but the physical meanings 
are same for both the cases. It is interesting to note that for super-radiance
to occur either eqn. (27a) or (27b) or both have to be satisfied at a radius $r \le r_c$
whatever be the $\sigma$. Thus, it is very obvious that for all $\sigma$ eqns. (27a.b)
do not satisfy, which is again in agreement with Chandrasekhar's calculation.  

\section*{III. Solution }
Here we will discuss about the solution of the reduced Maxwell equations
described in previous section. We mainly will concentrate on the solution
of eqns. (\ref{angtreq}), (\ref{rad1dec}), (\ref{rad2dec}) and will understand
about the dynamical behaviour of the spin-1 particles in Kerr geometry. 

\subsection*{III.A. Angular Solution}
From eqn. (\ref{radangf}), we see that for Schwarzschild black hole,
solution of angular set of equations is nothing but the standard spherical
harmonics $_s\!Y_{lm}(\theta,\phi)$ with integral azimuthal quantum
number $m$\cite{np,sudarsan}. In that occasion ${\cal L}^\dagger$ and
${\cal L}$ simply act as raising and lowering operator respectively of spin-s for
$_s\!Y_{lm}(\theta,\phi)$. Corresponding eigenvalue which is the separation
constant $\lh$ for spin-1 particles is appeared as 
$\lh^2=l(l+1)$\cite{sudarsan}, which is well known as the total angular momentum
of the system where $l$ is the angular momentum quantum number. 
For the case of spin-1/2 particles, Chakrabarti\cite{chakra}
found the angular solution in Kerr geometry in terms of the combination of 
different spherical harmonics and Clebsch-Gordan coefficients in perturbative 
method with $a\sigma$ as the perturbative parameter.
He also found the separation constant which was modified from that 
of Schwarzschild case in terms of Clebsch-Gordan coefficients. 
Here, our approach to find the angular solution is different from that
of Chakrabarti. Actually our method is easily possible to apply here even
for angular equation as particles are massless. Thus to get
angular solution we will solve eqn. (\ref{angtreq}) by IWKB 
method which was applied earlier for the solution of 
radial Dirac equation only. By this IWKB ({\it instantaneous WKB})
method we can get the analytical solution of the second order wave
equation which is described in detail in earlier papers\cite{mc99,mc00,m99}. 
By this method the transmission and reflection coefficients are calculated
at the each location. Here, the particle under consideration is moving
in a potential field varying with the location. As the potential barrier felt
by the particle is changing with space (see figures for potential plot), the
transmission and reflection amplitudes should be space-dependent. Thus from
the IWKB solution we can get the {\it local} values of the reflection and transmission
coefficient and the spatially complete solution
is possible to construct. One can have more clear feeling about the local
transmission and reflection coefficients if the potential barrier of the
system is thought to compose of a large number of square steps.
The size and number of the different steps are chosen in such a way that the overall
combination of all the steps follows the pattern of the actual potential
barrier\cite{cm00}. Now, one can solve this as a barrier problem where the barrier
under consideration contains a large number of square steps. The transmission
and reflection coefficients are calculated at each junction out of the transmission
in the previous junction, from what the meaning of local values of the transmission
and reflection coefficient is well-understood. In the earlier papers\cite{mc99,mc00}, in
the context of the solution of Dirac equation it was seen that in both the
ways (by IWKB method and the way when barrier is considered to compose by a 
large number of square steps) the solutions are matching perfectly. Here, throughout
the study we will concentrate on IWKB solution (not the barrier method as\cite{cm00})
to obtain an analytical result. 
The solution of IWKB method is valid if the wave number $k$ of the incoming 
photon at a location $x$ satisfies the condition $\frac{1}{k}\frac{dk}{dx}<<k$.
Thus, the applicability of IWKB method depends on the validity of the
WKB approximation method. 

In the eqn. (\ref{angtreq}), $\lh$ is already known following the calculation of  
Chakrabarti\cite{chakra}. Thus all the parameters are known in (\ref{angtreq}).  
As ${\cal J}_{\pm 1}$ varies in entire region and we are interested about the
dynamics of the waves in that region itself, from eqn. (\ref{angtreq}) 
we get the solution
\setcounter{equation}{27}
\begin{eqnarray}
\label{angsol}
S_{\pm 1}=\frac{{{\cal A}_a}_\pm}{\sqrt{{\cal J}_{\pm 1}}}e^{+iu_{a \pm}}+
\frac{{{\cal B}_a}_\pm}{\sqrt{{\cal J}_{\pm 1}}}e^{-iu_{a \pm}}
\end{eqnarray}
by IWKB method, where the coefficients ${{\cal A}_a}_\pm$ and 
${{\cal B}_a}_\pm$ are space-dependent related as ${{{\cal A}_a}_\pm}^2+
{{{\cal B}_a}_\pm}^2 ={\cal J}_{\pm 1}$\cite{mc99} because sum of the transmission
and reflection coefficients are always unity. The ikonals are 
defined as $\int {\cal J}_{\pm 1} dU = u_{a \pm}$. 

\begin{figure}
\vbox{
\vskip -0.5cm
\hskip 0.0cm
\centerline{
\psfig{figure=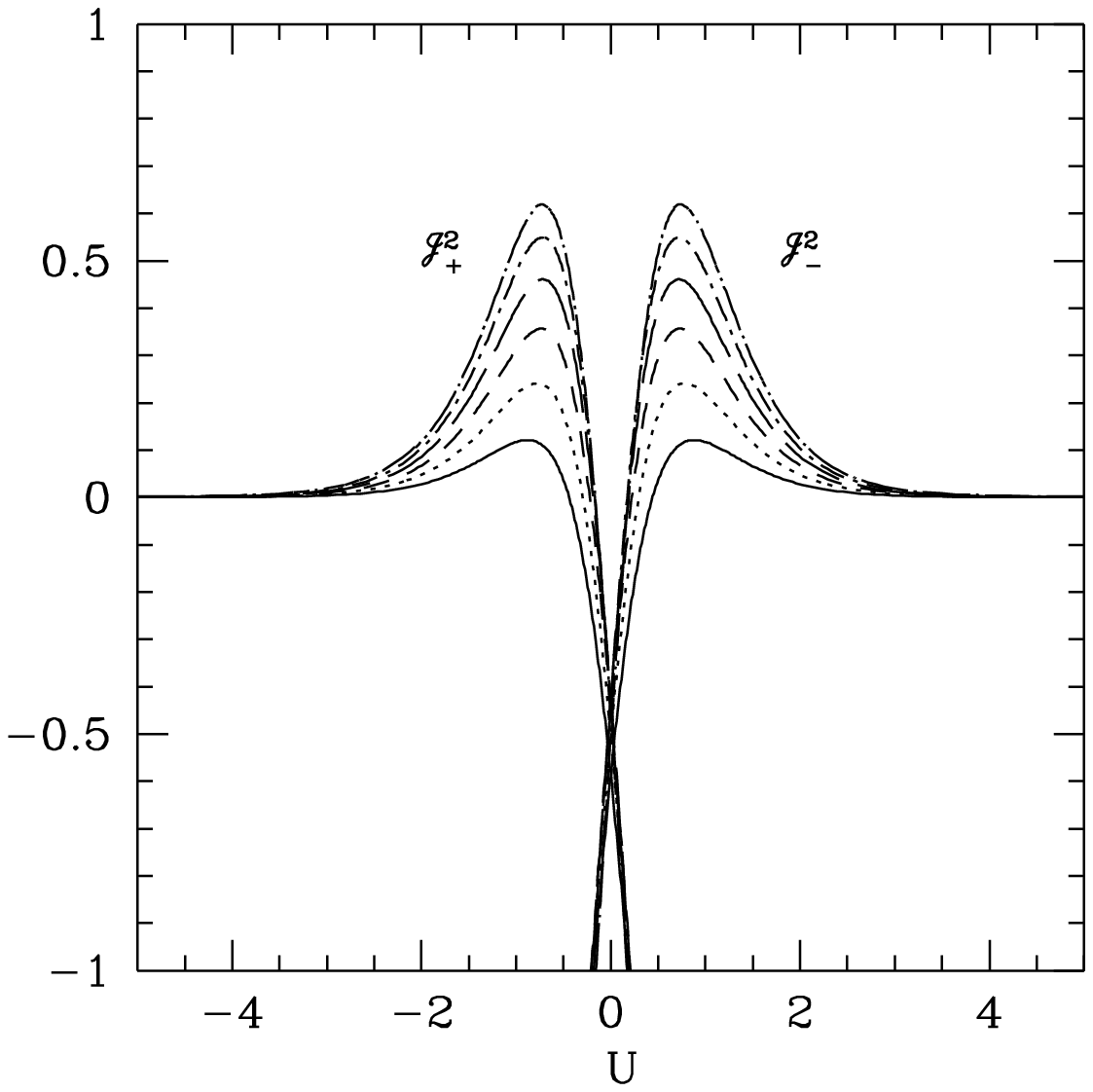,height=14truecm,width=14truecm,angle=0}}}
\vspace{-0.0cm}
\noindent {\small {\bf Fig. 1} : Variation of ${\cal J}_+^2$ and ${\cal J}_-^2$
as functions of $U$ for frequency $\sigma=0.5$ and Kerr parameter 
$a$= $0$ (solid curve), $0.2$ (dotted curve), $0.4$ (dashed curve), $0.6$
(long-dashed curve), $0.8$ (dot-dashed curve) and $1$ (dot-long-dashed curve). 
$a=0$ corresponds to Schwarzschild solution for which $\lh=2$. Following 
\cite{chakra} we get other $\lh$s as $\lh^2=1.704382, 1.417009, 1.136979, 0.863197,
0.594325$ for $a\sigma=0.1,0.2,0.3,0.4,0.5$ respectively; $l=1$, $m=-1$, $M=1$. 
}
\end{figure}

Now we will display the solutions for a few sets of physical parameter.
The physical parameters are chosen in such a manner that the interaction
of photon with black hole should be significant. Therefore, corresponding 
Compton wavelength should be of same order to the black hole radius, i.e.,
$\frac{1}{\sigma}\sim M+\sqrt{M^2-a^2}$ (as $\ch=G=c=1$), otherwise the
result is insignificant. Also we choose $M=1$ for all the figures.
Figure 1 shows the variation of ${\cal J}_+^2$ and ${\cal J}_-^2$
with respect to $U$ for different rotation parameters ($a=0\rightarrow 1$)
of the black hole for a particular frequency of the incident wave 
($\sigma=0.5$). It is seen that in a certain range both 
${\cal J}_+^2$ and ${\cal J}_-^2$ attain negative values and corresponding
ikonals become imaginary. Recalling Schr\"odinger wave equation, we can
say for that range of $U$ total energy of the incoming particle is
less than the potential energy of the system and particle is inside the
barrier. In the other range, total energy of the spin-1 particle 
(in that coordinate system) is greater than or equal to the potential energy 
of the system as ${\cal J}_\pm^2$ attains greater than or equal to zero 
respectively. In the case of ${\cal J}_+^2$, at around $\theta=\frac{\pi}{2}$
particle enters into the barrier and remains inside upto $\theta=\pi$.
On the other hand, behaviour of ${\cal J}_-^2$ is just opposite. In that
case particle energy becomes greater than or equal to the potential energy
of the system at around $\theta=\pi/2$ but for $\theta\le\pi/2$ the
particle remains inside the barrier. In this way, with the increment
of $\theta$ values, particle goes inside the barrier and then comes out
from it alternatively and corresponding $U$ runs from $-\infty$ to $\infty$
and $\infty$ to $-\infty$ alternatively. 
As the rotation of the black hole decreases curvature of the
space-time reduces, which reflects in the behaviour of ${\cal J}_\pm^2$ whose 
height is lowered down. The results with particular $a$ parameter but different
$\sigma$s will be same as Fig. 1 as it depends only on $a\sigma$ not
on the individual values of $a$ and $\sigma$.

Although the solution (\ref{angsol}) is looking like a wave 
solution containing incident and reflected parts but there is nothing like
that physically. This appears like that because of the method we choose and corresponding
transformation of variable $\theta$ to cartesian like variable $U$. Following
\cite{mc99}, \cite{mc00} and \cite{m99} we can find out the $\theta$-dependent 
expressions of ${\cal A}_{a_\pm}$ and ${\cal B}_a{_\pm}$ but there
is no relation to transmission and reflection wave amplitudes.
To calculate the $\theta$-dependent expressions of ${\cal A}_{a_\pm}$ and 
${\cal B}_a{_\pm}$ we need to impose the boundary conditions properly. In case
of ${\cal J}_+^2$, for $U\rightarrow -\infty$ (actually for $U\lsim -5$, see
Fig. 1) both ${\cal A}_{a_+}$ and
${\cal B}_{a_+}$ should be constant which are same as for pure WKB solution. 
But in the region $U\sim 0\rightarrow \infty$, as because the wave is inside
the potential barrier, ${\cal B}_{a_+}$ must be zero otherwise the solution
will diverge at $U\sim\infty$. In case of ${\cal J}_-^2$, the boundary
conditions for ${\cal A}_{a_-}$ and ${\cal B}_{a_-}$ should be appeared
just in an opposite way as is depicted from the behaviour
of ${\cal J}_-^2$ too. But it can be reminded that this solution is only valid for 
$\frac{1}{{\cal J}_\pm}\frac{d{\cal J}_\pm}{dU}<<{\cal J}_\pm$.

\subsection*{III.B. Radial Solution}

To find the solution of eqns. (\ref{rad1dec}) and (\ref{rad2dec}) we again
follow IWKB method where ${\cal K}_\pm$ a is function of $r$ 
and we are interested about the solution at the varying ${\cal K}_\pm$ region.
The solution is
\begin{eqnarray}
\label{radsol}
\psi_{\pm}=\frac{{{\cal A}_r}_\pm}{\sqrt{{\cal K}_{\pm}}}e^{+iu_{r \pm}}+
\frac{{{\cal B}_r}_\pm}{\sqrt{{\cal K}_{\pm}}}e^{-iu_{r \pm}},
\end{eqnarray}
where, the space dependent transmitted and reflected amplitudes are
related as ${{{\cal A}_r}_\pm}^2+{{{\cal B}_r}_\pm}^2={\cal K}_{\pm}$
and the ikonals are defined as $\int {\cal K}_{\pm} dV = u_{r \pm}$. In a
similar way as the case for angular solution sum of the 
space-dependent transmission and reflection coefficients are unity always.
Close to the black hole horizon
${{\cal B}_r}_\pm\hskip0.15cm{\rm should\hskip0.1cm be}\sim0$; 
because of high gravitative power there is virtually no outwards
reflection. Also, far away from the black hole, both ${{\cal A}_r}_\pm$ and 
${{\cal B}_r}_\pm$ are constant as potential does not vary there 
(see Fig. 2a,b). These are the essential boundary conditions to obtain
space-dependent solutions. Far away from the black hole, solutions will
merge to that of WKB solution. As matter comes closer, spatial variance
comes into the picture and IWKB method is needed to obtain an analytical
space-dependent solution.
This solution is valid for the entire range of $r$ and all values of 
$a$ (even for Schwarzschild
$a=0$ case). As for different physical parameter set, variations of
${\cal K}_{\pm}$ are different, corresponding space-dependent behaviour
of transmission and reflection coefficients will be different.

Here we display the solutions for a few physical parameter set for which 
the interaction between incoming photon and black hole is significant,
as explained in \S III.A.
Figure 2a shows the behaviour of the square of wave numbers ${\cal K}_+$ and ${\cal K}_-$
with respect to new variable $V$ for a particular frequency ($\sigma$) and
different angular momentum of the black hole ($a$). It is seen that
as $a$ value increases, curvature effect on the space-time increases
as a result corresponding strength of the wave number increases.
Similar features are seen in Fig. 2b where ${\cal K}_+^2$ and ${\cal K}_-^2$
are shown for a particular Kerr parameter ($a=0.5$) and different $\sigma$s.
As the frequency of the incident wave becomes higher, its energy becomes higher too,
as a result it needs to overcome the less strength of barrier. On the 
other hand with the decrement of frequency its energy decreases as well
as it feels higher potential barrier in the motion. For both the cases,
in entire region of $V$, ${\cal K}_\pm^2$ never be negative, which indicates
total energy of the incoming wave is always greater than or equal to the potential
energy of the system. It can be checked that for some parameter set,
${\cal K}_\pm^2$ becomes negative, which indicates the existence of 
the potential barrier whose peak is higher than the energy of the 
incoming particle as was seen in the case of Dirac particles in Kerr 
geometry\cite{mc00}.

\begin{figure}
\vbox{
\vskip -0.5cm
\hskip 0.0cm
\centerline{
\psfig{figure=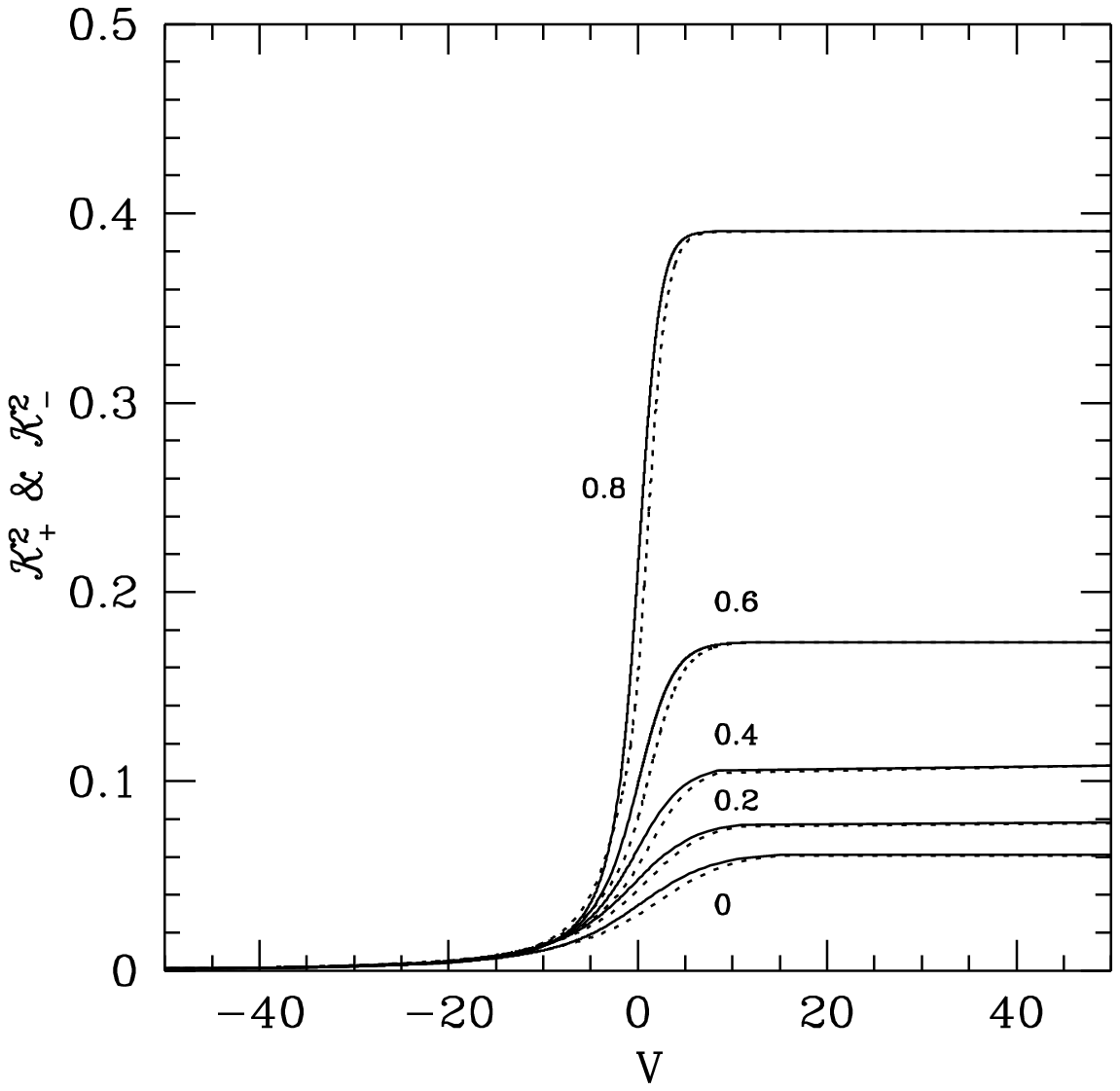,height=14truecm,width=14truecm,angle=0}}}
\vspace{-0.0cm}
\noindent {\small {\bf Fig. 2a} : Variation of ${\cal K}_+^2$ (solid curve) and 
${\cal K}_-^2$ (dotted curve) as functions of $V$ for a particular frequency 
$\sigma=0.5$ and different Kerr parameter $a$ as marked on each set of curve.
Corresponding values of $\lh$ for different plots are given in Fig. 1;
$l=1$, $m=-1$, $M=1$.
}
\end{figure}

\begin{figure}
\vbox{
\vskip -0.5cm
\hskip 0.0cm
\centerline{
\psfig{figure=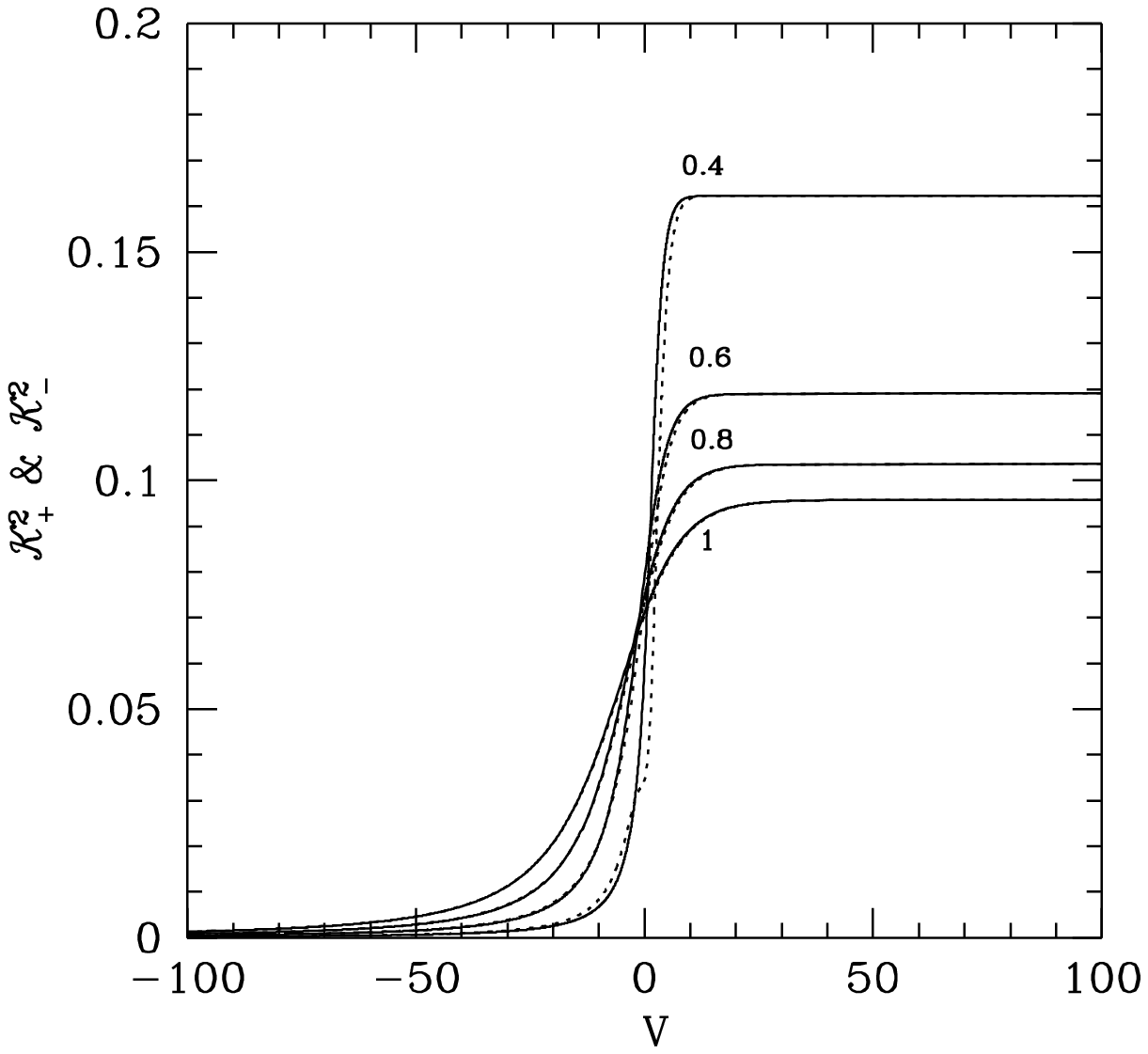,height=14truecm,width=14truecm,angle=0}}}
\vspace{-0.0cm}
\noindent {\small {\bf Fig. 2b} : Variation of ${\cal K}_+^2$ (solid curve)
and ${\cal K}_-^2$ (dotted curve) as functions of $V$ for a particular
Kerr parameter $a=0.5$ and different frequency of incident wave as marked 
on each set of curve.
Corresponding values of $\lh$ for different plots are given in Fig. 1;
$l=1$, $m=-1$, $M=1$.
}
\end{figure}

\begin{figure}
\vbox{
\vskip -0.5cm
\hskip 0.0cm
\centerline{
\psfig{figure=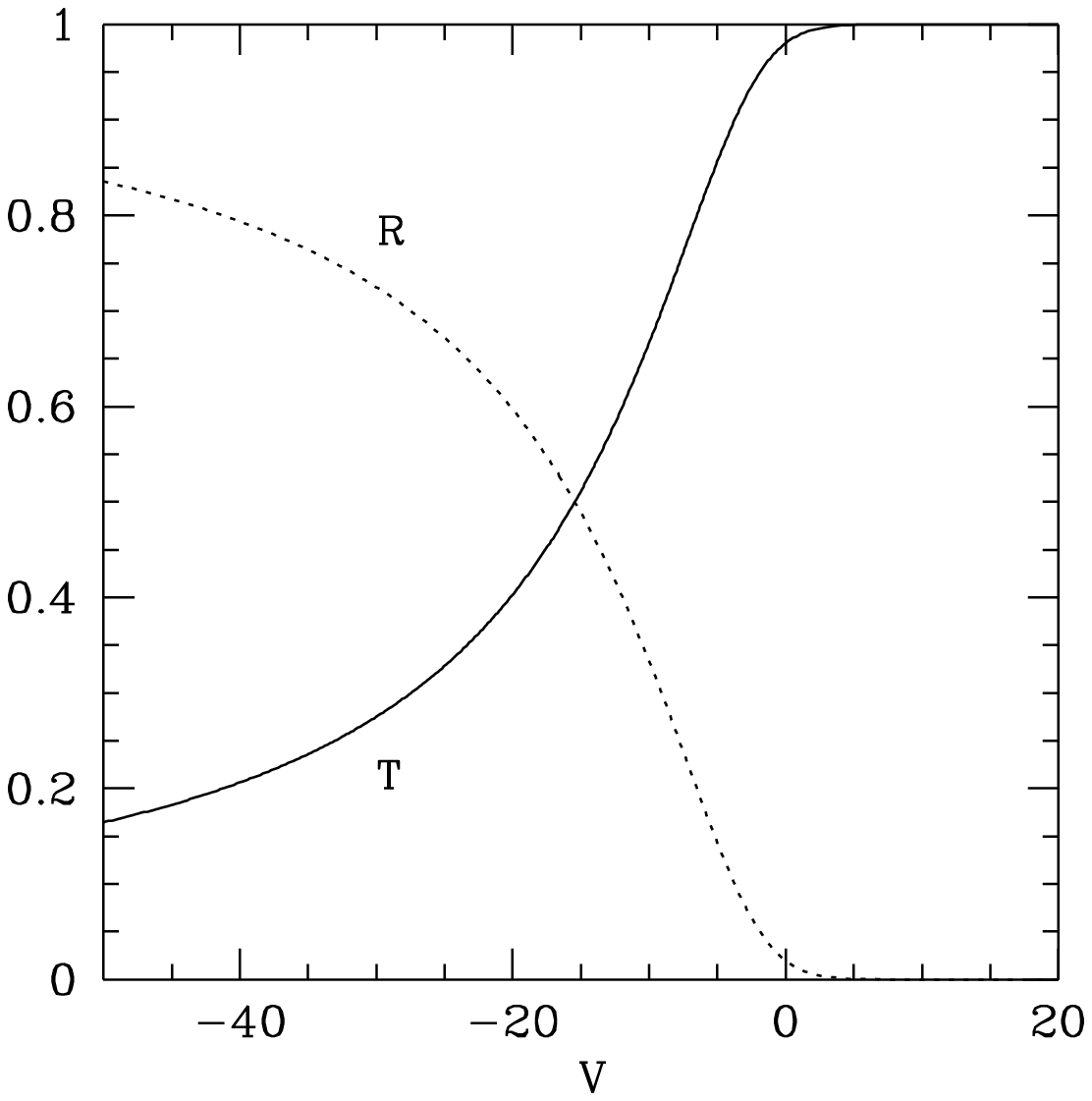,height=14truecm,width=14truecm,angle=0}}}
\vspace{-0.0cm}
\noindent {\small {\bf Fig. 2c} : Variation of transmission (solid curve) and
reflection (dotted curve) coefficients as functions of $V$ for $\sigma=0.5$,
$a=0.4$ and associated $\lh=1.190382$; $l=1$, $m=-1$, $M=1$.}
\end{figure}

Figure 2c shows the space-dependent reflection and transmission 
coefficients for the frequency of particle
$\sigma=0.5$ and angular momentum of the compact object $a=0.4$. Following
\cite{chakra} we get corresponding $\lh=1.190382$. In Fig. 2a, the corresponding
behavior of wave number is shown. Far away
from the black hole ($V\rightarrow -\infty$) small ${\cal K}_+$ indicates higher
barrier hight and high rate of reflection. As the wave comes closer
to the black hole ($V\rightarrow\infty$) potential barrier decreases
as well as wave number ${\cal K}_+$ increases which results particle becomes
free or almost free. So in that region transmission is 100\% or almost
100\%. At the other locations, these transmission and
reflection coefficients are varying according to the behaviour of potential
as well as ${\cal K}_+$. Again, this solution is only valid as the condition
$\frac{1}{{\cal K}_\pm}\frac{d{\cal K}_\pm}{dV}<<{\cal K}_\pm$ satisfies.

It can be mentioned that this method of solution is also applicable to a
range of parameters which demonstrates super-radiance where potential 
diverges at a certain location $r>r_+$. In early, this divergent nature of potential
was depicted in case of fermions\cite{mc00}, though for the case of fermion
super-radiance does not occur. In the present case, if we choose for example,
$a=0.5$, $\sigma=0.2$ and corresponding $\lh=1.305519$\cite{chakra}, $m=-1$,
it is found that potential as well as ${\cal K}_\pm^2$ diverge at a 
certain location $r>r_+$ and $V-r$ relation becomes multivalued. For this parameter set 
super-radiance is expected to occur. Similarly, different other parameter set can be 
chosen where potentials as well as wave numbers diverge. 

\section*{VI. conclusions}

Here we have studied the solution of Maxwell's equation in curved space-time
particularly in Kerr geometry analytically. By the solution we can get the 
dynamical behaviour of spin-1 particles around black holes. We started with
Maxwell equations in curved space-time especially in Kerr geometry then separated
it into radial and angular parts. Furthermore, the radial and angular variables
are given transformation in such a way that the equations are reduced to
one dimensional wave equations in cartesian like coordinate system 
which is easier to attack. Following the earlier
works by the author radial and angular solutions of Maxwell equations are
obtained which we had only asymptotic knowledge\cite{chandra83} so far. 
Here we give the spatially complete analytical solution which is helpful for further studies
those are already mentioned in the Introduction. 
Here the incident and reflected 
amplitudes are not constant, which vary according to the variation of the
potential. Thus the reflection and transmission rate of the incident photon
are space-dependent. On the occasion for the solution of Dirac equation
in a black space-time author initiated a technique namely IWKB method
to obtain the space-dependent behaviour of the transmission and reflection 
coefficient and corresponding spatially complete solution, then it was verified 
with the numerical results\cite{mc99,mc00,cm00,m99}. Here we have used same IWKB
method to obtain solutions of Maxwell equations. The detail technique, i.e.,
how to obtain the space-dependent analytical expressions for transmission and
reflection amplitudes are not repeated here. Here, one of the main emphasis is 
to construct an analytical complete solution of the electromagnetic wave in a 
curved space-time, according to my knowledge which was unavailable so far in the 
literature although the asymptotic as well as numerical solutions were available. 
Once we have the analytical solution, we can apply it obtain the Feynman Green 
function, second quantization in a specific metric, here for Kerr metric.
The technique applied here is quantum mechanical where $S_{\pm 1}$ and $\psi_{\pm}$
are wave functions and the corresponding amplitudes of the wave are number.
Once we have this knowledge of solution we can make second quantization in
curved space-time where $S_{\pm 1}$ and $\psi_{\pm}$ are no longer wavefunctions
but fields and the corresponding amplitudes become operator. Also in the context
of Hawking radiation, to check black hole's stability and its perturbation,
to find the quasi-normal modes of the black hole this solution is useful, as
this is valid at close to the black hole as well as away from it.   

Also we have studied here how the electromagnetic wave face different potential 
barrier for different physical parameter in a black hole space-time. As the 
radial functions $R_{\pm 1}$ are given a certain choice of transformation, 
the potential in the final second order wave equation becomes real (unlike
the earlier works like \cite{chandra83} etc.) which is now easier to study. 
According to that we concentrated on to study the transmission and
reflection coefficients of the incoming wave.
It is seen that depending on the frequency of the incoming electromagnetic
wave and the angular momentum of the central object potential as well as
wave number of the system change. Thus, it is concluded that the black
hole can distinguish the spin-1 particles with different physical parameters.
Earlier it was shown\cite{mc99} by the author that 
black hole can distinguish Dirac particles of different frequency and mass and act 
as mass-spectrograph. Now globally we can say that black hole can act as
mass-spectrograph irrespective of the spin, frequency and mass of the particle
(although in case of electromagnetic wave there is no question of rest mass). 
Furthermore, the coupling between the spin of black hole and that of particle
parts important role, as a result, particle of the particular physical parameter will face
different potential barrier for different black holes. Also the same particle with
spin-up and spin-down in case of a particular black hole will face totally
different potential. Thus we can conclude that the gravitational field is very
sensitive on the spin of the particle.

Another important issue is the super-radiance. It was seen\cite{mc00} that
for Dirac particle there is no existence of super-radiance. In case of
spin-1 particle super-radiance is very obvious feature as was conjectured
by Chandrasekhar earlier\cite{chandra83}. But the existence of super-radiance
disappears as particle flips in its spin. For the same sign of spin of the
black hole and particle (here $a$ and $m$ respectively) with positive energy of the
particle super-radiance does not occur, but, if any one of them flips in spin
super-radiance may appear for a particular range of frequency.

\vskip0.5cm
\noindent{\large\bf Acknowledgment}\\
I would like to express my gratitude to Professor A. R. Prasanna for
carefully reading the manuscript and continuous encouragement. 

I also would like to dedicate this work to my late wife Aruna Mukhopadhyay,
who was passed away on 15th March, 2001. Because of her continuous 
encouragement my research life was very easy earlier. I express my deep 
condolence for her.

\end{document}